\documentclass{emulateapj}
\def\kh{KH~15D~}
\def\kms{km~s$^{-1}$}

\def\en{0.27}

\def\csn{1.5}
\def\cs{$\sqrt{\chi_\nu^2} = \csn$}

\def\rms{0.38}
\def\kn{8.15}

\def\vrn{6.9}

\def\pn{48.38}

\def\nobs{16}

\def\msun{M$_\odot$}

\def\ppn{10.7}  %peak-to-peak radial velocity variations

\begin{document}

\title{KH~15D: A Spectroscopic Binary~$^1$}
         
\author{ John Asher Johnson\altaffilmark{2}, 
	Geoffrey W. Marcy\altaffilmark{2},
	Catrina M.  Hamilton\altaffilmark{3,4},
	William Herbst\altaffilmark{5},
	Christopher M. Johns-Krull\altaffilmark{3,4}
 }
	 
\altaffiltext{1}{Based on observations obtained at the W. M. Keck
Observatory, which is operated jointly by the University of California and
the California Institute of Technology; Las Campanas Observatory of the
Carnegie Institution with the Magellan II Clay telescope ; and
McDonald Observatory of the University of Texas at Austin.} 
\altaffiltext{2}{Department of Astronomy, University of California,
Mail Code 3411, Berkeley, CA 94720}
\altaffiltext{3}{Visiting astronomer, McDonald Observatory which is operated by
The University of Texas at Austin.}
\altaffiltext{4}{Department of Physics and Astronomy, Rice University,
Houston, TX 77005.}
\altaffiltext{5}{Van Vleck Observatory, Wesleyan University, Middletown, CT 06459}

\begin{abstract}
We present the results of a high-resolution spectroscopic
monitoring program of the eclipsing pre--main-sequence star KH~15D
that reveal it to be a single-line spectroscopic binary. We
find that the best-fit Keplerian 
model has a period $P = \pn$ days, which is nearly identical to the
photometric period. Thus, we find the best 
explanation for the periodic dimming of KH 15D is that the binary 
motion carries the currently visible star alternately above
and below the edge of an obscuring cloud. The data are consistent with
the models involving an inclined circumstellar disk, as recently proposed by
Winn et al. (2004) and Chiang \& Murray-Clay (2004). We show that the
mass ratio expected from models of PMS evolution, together with the
mass constraints for the visible star, restrict the orbital
eccentricity to $0.68 \le e \le 0.80$ and the mass function to $0.125
\le F_M/\sin^3i \le 0.5$~\msun.  
\\
\\
\noindent {\em Subject Headings:} 
	      stars: pre--main-sequence --- 
              stars: individual (KH~15D) ---
	      binaries: spectroscopic ---
	      techniques: radial velocities ---
	      circumstellar matter
\end{abstract}

\section{Introduction}
\kh is a K6-K7 pre--main-sequence star 
that exhibits dramatic photometric variability
\citep{kearns98}. Every 48 days, the star's
brightness dims by 
3.5 magnitudes and remains in this faint 
state for nearly half of the photometric period. These deep brightness
minima are accompanied by a slight blueing of the star's color indices
\citep[][hereafter He02]{herbst02},
little or no change in spectral type 
\citep{hamilton01} and an increase 
in linear polarization \citep{agol04}. This implies that the
star is completely eclipsed by an optically thick, extended collection of
dust grains, possibly in the form of a circumstellar
disk. If this is the case, the serendipitous alignment of the \kh
star/disk system may provide insights into the
evolution of young stars and their interactions with their
circumstellar environments. 

While \kh has periodic eclipses, it cannot be an ordinary eclipsing
binary because of the long duration of minimum light.
Recent theories postulate that there is, nonetheless, a 
currently unseen binary companion to the visible star.
In a study of archival photographic plates,
\citet[][hereafter JW04]{johnson04} discovered that the historical
light curve of \kh is similar to the modern light curve but
appears to be diluted by light from a second star. Motivated by these
findings, \citet[][hereafter W04]{winn04} constructed a model composed
of a binary system with the orbital plane inclined with respect to 
the edge of an optically thick screen. As the two stars orbit one
another, the reflex motion carries one star alternately above and below
the edge of an opaque screen, causing the eclipses.  
The long-term evolution of the light curve is reproduced by allowing the screen
to move slowly across the binary orbit---a feature of the model that
led the authors to envision the screen as a precessing,
circumbinary disk. A similar model 
is proposed by \citet[][hereafter CM04]{chiang04} who also envision
the opaque screen as an inclined, precessing circumbinary disk, or
``ring,'' with an inner edge truncated by tidal interactions with the
binary and an outer 
edge possibly shepherded by a planetary companion. Both models provide
explanations of the unique features of KH~15D's light curve. The W04
model makes quantitative predictions about the 
orbital parameters of the binary system, while the CM04 model provides
a physical description of the circumbinary ring.
\begin{deluxetable*}{rlccccc}
\tablecaption{Spectroscopic Observations of \kh \label{observations}}
\tablewidth{0pt}
\tablehead{
\colhead{UT Date} &
\colhead{Telescope/} &
\colhead{$\Delta\lambda$ [\AA]} &
\colhead{R} &
\colhead{J.D.$- 2.4\times 10^6$} &
\colhead{$v_r$ [km/s]} \\
\colhead{} &
\colhead{Instrument} &
\colhead{} &
\colhead{} &
\colhead{} &
\colhead{} &
}
\startdata
25 Oct 2002 & Keck/HIRES & 3900-6200 & 70,000 & 52572.574 & $1.7 \pm 0.2$\tablenotemark{a}\\
28 Oct 2002 & Keck/HIRES & 3900-6200 & 70,000 & 52575.511 & $3.0 \pm 0.3$\tablenotemark{a}\\
16 Dec 2002 & Magellan/MIKE  & 4500-6800 & 45,000 & 52624.738 & $3.1 \pm 0.2$\tablenotemark{b}\\
14 Jan 2003 & Keck/HIRES & 3900-6200 & 70,000 & 52653.468 & $3.3 \pm 0.2$\tablenotemark{a}\\
8 Feb 2003  & Keck/HIRES & 4200-6600 & 70,000 & 52678.400 & $9.0 \pm 0.5$\tablenotemark{b}\\
9 Feb 2003  & Keck/HIRES & 4200-6600 & 70,000 & 52679.410 & $11.5 \pm 0.5$\tablenotemark{b}\\
9 Mar 2003  & Magellan/MIKE  & 4900-8700 & 22,000 & 52707.511 & $0.8 \pm 0.2$\tablenotemark{b}\\
3 Nov 2003  & Keck/HIRES & 4200-6600 & 30,000 & 52946.584 & $1.2 \pm 0.4$\tablenotemark{b}\\
4 Nov 2003  & Keck/HIRES & 4200-6600 & 30,000 & 52947.594 & $1.8 \pm 0.3$\tablenotemark{b}\\
 4 Jan 2004 & McDonald/CE & 5600-6900 & 30,000 & 53008.762 & $1.4 \pm 0.6$\tablenotemark{b}\\
 5 Jan 2004 & McDonald/CE & 5600-6900 & 30,000 & 53009.852 & $1.7 \pm 0.4$\tablenotemark{b}\\
10 Jan 2004 & Keck/HIRES & 3900-6200 & 70,000 & 53014.320 & $5.4 \pm 0.3$\tablenotemark{a}\\
10 Jan 2004 & McDonald/CE & 5600-6900 & 30,000 & 53014.760 & $5.8 \pm 0.4$\tablenotemark{b}\\
11 Jan 2004 & Keck/HIRES & 3900-6200 & 70,000 & 53015.506 & $7.0 \pm 0.2$\tablenotemark{b}\\
9 Feb 2004  & Keck/HIRES & 3900-6200 & 70,000 & 53044.834 & $1.8 \pm 0.2$\tablenotemark{a}\\
10 Feb 2004 & Keck/HIRES & 3900-6200 & 70,000 & 53045.828 & $1.3 \pm 0.2$\tablenotemark{a}
\enddata
\tablenotetext{a}{Multiple reference stars were
used to calculate
the radial velocity. The reported velocity is the mean velocity
measured from the various target-reference star pairs. The estimated
uncertainty is the standard deviation of the mean velocity. See Table
\ref{multiobs} for a listing of individual reference stars and radial
velocity measurements for each target-reference star pair.}
\tablenotetext{b}{Only one reference star was used in the radial velocity
measurement. The reported velocity is the mean velocity from all
echelle orders. The estimated uncertainty is the standard
deviation of the mean velocity.}
\end{deluxetable*}

A fundamental question that has not yet been answered is
whether \kh~is a single or multiple stellar system.
He02 \citep[see also][hereafter Ha03]{hamilton03} first
searched for evidence of orbital companions using high-resolution VLT
spectra and reported a  
radial velocity change of $+3.3 \pm 0.6$ km s$^{-1}$ over two widely spaced
epochs. However, one 
of the measurements was made during egress when there was 
a strong 
possibility of contamination from scattered light or 
line profile distortions from a sharp-edged cloud only partially
occulting the stellar disk.

Over the past two years we have conducted a high-resolution, multi-site
spectroscopic monitoring campaign to determine 
whether or not \kh exhibits orbital motion indicative of a
multiple system. Here we present
the results of our study, which show that \kh undergoes
significant
radial velocity variations. The variations are consistent with a
binary companion with an orbital period equal to the
48-day photometric period. In \S \ref{data} we summarize our observations
and reduction procedures. The radial velocity measurements and
best-fit Keplerian orbital parameters are presented in \S
\ref{analysis}. We conclude in \S \ref{discussion} with a discussion
of our findings and implications for existing models of the \kh eclipse
mechanism.

\section{Data}
\label{data}
During the 2002-2003 and 2003-2004 observing seasons we made
\nobs~observations of \kh at maximum light including 11 
spectra with the Keck 10-meter telescope and HIRES echelle
spectrometer; 2 spectra with the 6.5-meter Magellan II (Clay) telescope and
the MIKE echelle spectrometer; and 3 spectra with the 2.1-meter Otto Struve
Telescope at McDonald Observatory with the Sandiford Cassegrain Echelle
Spectrometer (CE). The observations are summarized in Table
\ref{observations}. Additional Keck/HIRES spectra were obtained during
minimum light. However, since it is unlikely that the star's photosphere is
visible through the obscuring material, we decided to exclude from our
analysis spectra that were obtained within 10 days of mid-eclipse.
In addition to our own measurements, we also include 
the out-of-eclipse radial velocity measurement, $v_r = 9.0 \pm
0.2$~\kms, reported by (Ha03) based on their VLT/UVES spectra.

Most of the Keck/HIRES observations were made as part
of the California \& Carnegie Planet Search.\footnote{{\tt
http://www.exoplanets.org}} For these observations, the
relatively faint apparent magnitude of KH~15D
($V = 16$) at maximum light precluded the
the use of the iodine cell to establish a wavelength scale. The
cell was therefore removed from the light path during observations of
KH~15D to increase the throughput of the spectrometer. The
raw CCD frames from all telescopes were reduced using reduction
packages written in IDL.  
The details of the reduction procedures are fundamentally identical to 
the algorithm presented by \citet{valenti94}. After bias-subtraction,
each echelle frame is divided by a normalized median flat-field image.
Order definition is performed using a bright star or flat-field
exposure, and scattered light is removed by fitting a two-dimensional
B-spline to the 
inter-order regions and interpolating across each spectral order. After
the scattered-light is subtracted, each order is
rectified, sky-subtracted and summed in the cross-dispersion
direction to form the final one-dimensional spectrum. 

Instead of a summation in the cross-dispersion direction, the 
rectified orders of the McDonald spectra are reduced to
one-dimensional spectra using
the optimal extraction algorithm described by \citet{hinkle00}. 
In the case of the MIKE reductions, the standard code is modified to
correct for 
the tilt of the spectrometer entrance slit with respect to the CCD
columns. The correction of the slit tilt is necessary
because the sky subtraction algorithm we employ requires that the projected
slit image lie parallel to the detector columns.

The radial velocity of \kh relative to the Solar System barycenter is
measured from each spectral observation by means of a cross-correlation
analysis. For the spectra of \kh obtained as part of the Planet 
Search observing program, the program stars observed on each night
provided an extensive selection of reference stars with known
barycentric radial velocities listed in \citet{nidever02}. We selected
reference stars that were observed
within 30 minutes of \kh and with spectral types
ranging from M0 to G5. Since the Planet Search
target stars are observed through an iodine cell, orders containing
iodine absorption lines are avoided in the analysis. For 
\kh observations obtained as part of programs other than the Planet
Search, a single K-type reference star observed either before
or after \kh is used as the reference star. 

\begin{deluxetable}{cccc}[!h]
\tablecaption{Absolute Radial Velocity Measurements of \kh
From Each Reference Star \label{multiobs}}
\tablewidth{0pt}
\tablehead{
\colhead{J.D.$- 2.4\times 10^6$} &
\colhead{Reference} &
\colhead{$v_r$ [km/s]\tablenotemark{a}} \\
\colhead{} &
\colhead{Star HD\#} &
\colhead{} &
}
\startdata
52572.574 & 40979 & $1.6\pm0.1$\\
          & 56274 & $2.1\pm0.1$\\
          & 52265 & $2.1\pm0.08$\\
          & 49674 & $1.3\pm0.09$\\
          & 66171 & $1.2\pm0.2$\\
52575.511 & 31966 & $2.9 \pm 0.2$\\
          & 36395 & $4.0\pm0.08$\\
          & 47157 & $3.2\pm0.2$\\
          & 37124 & $2.2\pm0.1$\\
          & 40979 & $2.7\pm0.3$\\
52624.738 & 10700 & $3.1 \pm 0.2$\\
52653.468 & 31966 & $3.6 \pm 0.1$\\
          & 42618 & $2.7\pm0.1$\\
          & 47157 & $3.1\pm0.07$\\
          & 99109 & $3.6\pm0.09$\\
          & 99492 & $3.4\pm0.1$\\
52678.400 & 36006 & $9.0 \pm 0.5$\\
52679.410 & 36006 & $11.5 \pm 0.5$\\
52707.511 & 36006 & $0.8 \pm 0.2$\\
52946.584 & 36006 & $1.2 \pm 0.4$\\
52947.594 & 36006 & $1.8 \pm 0.3$\\
53008.762 & 36006 & $1.4 \pm 0.6$\\
53009.852 & 17660 & $1.7 \pm 0.4$\\
53014.320 & 45350 & $5.1\pm 0.2$\\
          & 18830 & $5.6\pm0.3$\\
53014.760 & 17660 & $5.8 \pm 0.4$\\
53015.506 & 45350 & $7.0 \pm 0.2$\\
53044.834 & 50499 & $2.0 \pm 0.2$\\
          & 46375 & $1.7\pm0.1$\\
          & 50281 & $1.8\pm0.2$\\
          & 46375 & $1.9\pm0.2$\\
53045.828 & 50499 & $1.6 \pm 0.1$\\
          & 76909 & $1.3\pm0.1$\\
          & 73667 & $0.9\pm0.1$\\
          & 72673 & $1.2\pm0.1$\\
\enddata
\tablenotetext{a}{Velocities are the mean value from all
echelle orders used in the cross-correlation analysis. Estimated
uncertainties are the standard deviation of the mean velocity from
all echelle orders.}
\end{deluxetable}

The cross-correlation of each \kh spectrum with respect to the
reference spectra is performed using custom routines written in IDL. The
procedure involves first rebinning each one-dimensional spectral 
order onto a new wavelength scale that is linear in $\log\lambda$.
This ensures that each pixel in the rebinned spectrum represents a
velocity interval that is uniform over the entire spectral order
\citep{tonry79}. Regions containing telluric lines, strong emission
features, and CCD defects are masked out and each spectral order is
cross-correlated with respect to the corresponding order of the reference
spectrum. Each spectral order thus yields an independent measurement
of the radial velocity of \kh relative to the reference star. The average
of the ensemble set of velocities from all orders is 
then adopted as the relative radial velocity of \kh for a given epoch.

The relative radial velocities from each night are converted into
absolute barycentric radial velocities using the relation

\begin{equation}
v_{rad} = \Delta v + (BC_{kh} - BC_{ref}) + v_{ref}.
\label{vrad}
\end{equation}

\epsscale{1.2}
\begin{figure}[!t]
\plotone{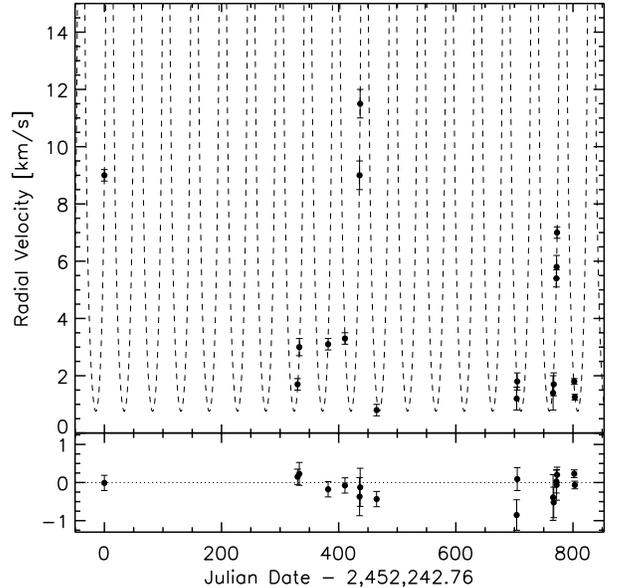}
\caption {The radial velocity of \kh as a function of
time. The dashed line is the best-fit Keplerian orbit with the
eccentricity fixed at $e = 0.74$---the mean value allowed by our orbit
constraints (see \S \ref{constrain}). }
\label{vel_vs_time}
\end{figure}

\epsscale{0.9}
\begin{figure*}[!t]
\plotone{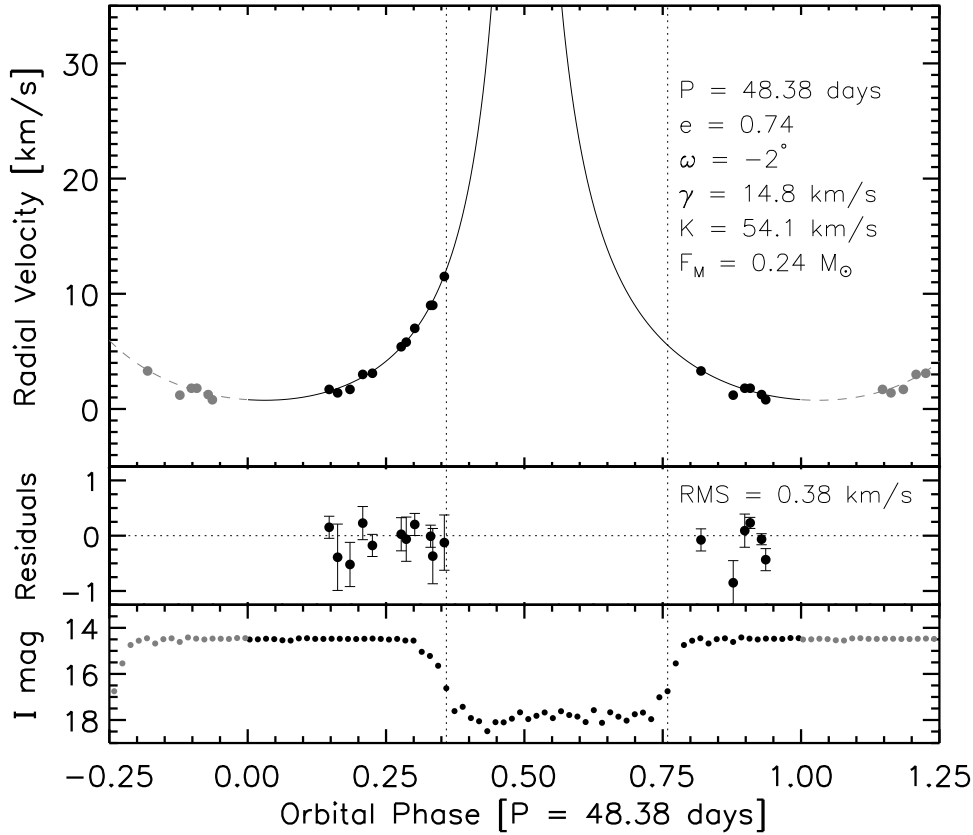}
\caption {The radial velocity of \kh as a function of orbital phase. The
solid line is the best-fit Keplerian orbit with the eccentricity fixed
at $e = 0.74$---the mean value allowed by our orbit constraints (see
\S \ref{constrain}). The
fit has reduced chi-squared \cs. The vertical dotted lines represent the
approximate phases of ingress (left) and egress (right) based on the
He02 ephemeris. The bottom panel shows the 2001-2002 {\em I}-band light
curve. The phased photometric measurements have been binned for
clarity using phase intervals of $\Delta \phi = 0.015$.}
\label{ThePlot}
\end{figure*}

\noindent In Eqn.~(\ref{vrad}), $\Delta v$ is the relative velocity 
from the cross-correlation analysis; $BC_{kh}$ and $BC_{ref}$ are 
the barycentric corrections for \kh and reference star,
respectively; and $v_{ref}$ is the absolute barycentric radial
velocity of the reference star 
as listed in \citet{nidever02}. The mean measured velocities from each
epoch are listed in the final column of Table
\ref{observations}. Table \ref{multiobs} lists the velocities obtained
from our cross-correlation analysis for each individual reference star
on each night.  

For nights when only one reference spectrum is
available, the uncertainty is estimated using the standard
deviation of the mean radial velocity measured from all
echelle orders. For nights with multiple reference spectra, the
standard deviation of the velocities computed from each
target-reference pair is adopted as the uncertainty. This latter
method of error estimation typically yields a larger value than the
order-to-order scatter seen for an individual observation ($0.2 \le
\sigma_v \le 0.6$~\kms compared to $0.1 \le \sigma_v \le 0.3$). The larger
scatter in velocity measurements among the target-reference star
combinations is likely due to external systematic effects such as
small shifts in the position of the CCD throughout the night caused
by thermal relaxation of the detector mount. Thus, the standard
deviation measured from the echelle orders in a single observation
likely underestimates the true uncertainty by approximately a factor
of two.

\section{Results}
\label{analysis}

Figure \ref{vel_vs_time} illustrates how the radial velocity of
\kh varies temporally out of eclipse over a range of
\ppn~km~s$^{-1}$. 
These data indicate that there must be an unseen
star in the system, as had previously been inferred from the analysis
of the historical light curve (JW04). Hereafter we will adopt the
naming convention of W04 and refer to the
currently visible star as A and the hidden companion as B. 

\subsection{Orbit Solution}
\label{orbit}

Using a nonlinear least-squares algorithm, we found a best-fit model orbit
with a period $P = \pn$~days, eccentricity $e \ge 0.27$
and velocity semiamplitude $K \ge \kn$~\kms. 
The other orbit parameters are listed in Table 
\ref{parameters}. Figure \ref{ThePlot} shows a plot of radial velocity
versus orbital phase for $P = \pn$ days. The rms scatter 
of the fit residuals is \rms~\kms and the reduced \cs. The vertical
lines at $\phi = 0.33$ and $\phi = 0.73$ denote the approximate phases
of ingress and egress, respectively, based on the He02 ephemeris. We
have no radial velocity measurements between these phases because of
the eclipse of star A. 

We find that the orbit solution is not well constrained due to the lack
of data near periapse, which allows the velocity semi-amplitude of the
orbit solution to compensate for changes in the eccentricity. As such, we were
able to obtain reasonable fits by fixing the eccentricity at values
$e > 0.27$, with each solution yielding different
values of the orbit parameters and values of $\sqrt{\chi_\nu^2}$ 
that are equivalent at the 97.5\% confidence level (based on the 9
degrees of freedom in the fit). 

\begin{figure}[!t]
\plotone{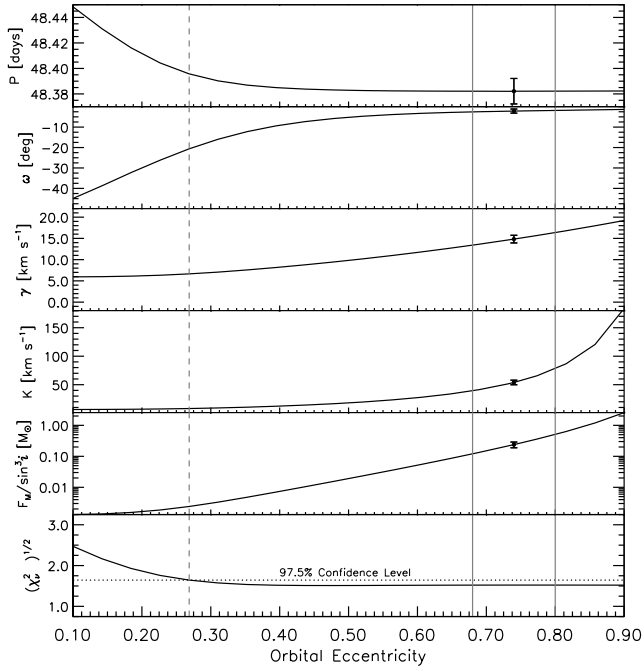}
\caption {The dependence of the best-fit orbit parameters (top 5
panels) and reduced chi-squared (lower panel) on the the orbital
eccentricity. The ordinate values are generated by fixing the
orbital eccentricity in the least-squares fit to the radial velocity
data. The vertical dashed line indicates the lower limits placed on
the fit parameters based on the 97.5\% confidence level for
$\sqrt{\chi_\nu^2}$ with 9 degrees of freedom. The solid vertical
lines denote the upper and lower limits placed on the eccentricity as
discussed in \S \ref{constrain} (see also Figure
\ref{mvse}), and the fits generated for $e = 0.74$ are denoted by solid
dots with error bars.}  
\label{versuse}
\end{figure}

Figure \ref{versuse} shows how $\sqrt{\chi_\nu^2}$ and the orbit parameters
vary as a function of eccentricity. While the period varies little
over a wide range of eccentricities, it is apparent that the radial
velocity data alone provide poor constraints for the other orbit
parameters. However, the implied 
mass ratio from other \kh observations, together with the measured mass
limits of star A, can be used to place limits on the orbital eccentricity, as
we now show. 

\epsscale{1.25}

\subsection{Orbit Parameter Constraints}
\label{constrain}

JW04 determined from photometric measurements of 
archival plates that the out-of-eclipse magnitude of \kh
was 0.9 mag brighter (at $I$-band) 40 years ago compared to the 
modern bright state. 
The two most probable explanations for the brighter
apparent magnitude in the past are that either both stars
were visible, or that star B
alone was visible. If both stars were visible, then $L_B/L_A =
1.3$. If only B were visible, then $L_B/L_A = 2.3$. In either case,
the condition $L_B/L_A > 1$ must hold.  

Multi-color photometric measurements obtained during minimum light
show a slight blueing of the color indices (Ha03)
compared to the colors at maximum light. Similarly,
\citet{agol04} measure slightly bluer colors during eclipse from their
low-resolution spectropolarimetric observations compared to their 
out-of-eclipse observations. One possibility for the bluer color
indices during eclipse is that the scattered light is dominated by Rayleigh
scattering. However, He02 show that there is no reddening
of the light from star A during ingress and egress. This suggests that
the opacity of the occulting material is wavelength--independent and
that the bluer colors are due to a bluer object. Therefore the
temperature of B must be hotter than the temperature of A,
assuming both stars contribute nearly equally to the scattered 
component of the light observed during the eclipse of A. However,
since the colors are only $\sim 0.1$~mag bluer at minimum light 
(He02), $T_B$ cannot be much larger than $T_A$.

For most low-mass ($M_* < 1.0$~\msun) pre--main-sequence evolutionary models
\citep[e.g][]{chabrier97,dantona97} the stellar mass 
is monotonic with both luminosity and temperature. For coeval
stars on their Hayashi tracks, $T_B \gtrsim T_A$ and $L_B/L_A > 1$ imply
that $M_B \gtrsim M_A$ or, in terms of the mass ratio, $M_A/M_B
\lesssim 1$. 

The mass function of a Keplerian orbit can be expressed as

\begin{equation}
F_M(e,K,P) = \frac{K^3 P(1-e^2)^\frac{3}{2}\sin^3i}{2\pi G} = \frac{M_B^3\sin^3i}{(M_A + M_B)^2},
\label{massfunc}
\end{equation}

\noindent where $P$ is the orbital period and $K$ is the velocity
semi-amplitude. Solving Eqn.~(\ref{massfunc}) for $M_A$ yields

\begin{equation}
%M_A(e, R) = \frac{M_A}{M_B}\left(1 + \frac{M_A}{M_B}\right)^2
%F(e,K,P).
M_A(e, M_A/M_B) = \frac{M_A}{M_B} 
\left(1 + \frac{M_A}{M_B}\right)^2 \left[\frac{F_M(e,K,P)}{\sin^3i}\right].
\label{ma}
\end{equation}

\epsscale{1.3}
\begin{figure}[!b]
\plotone{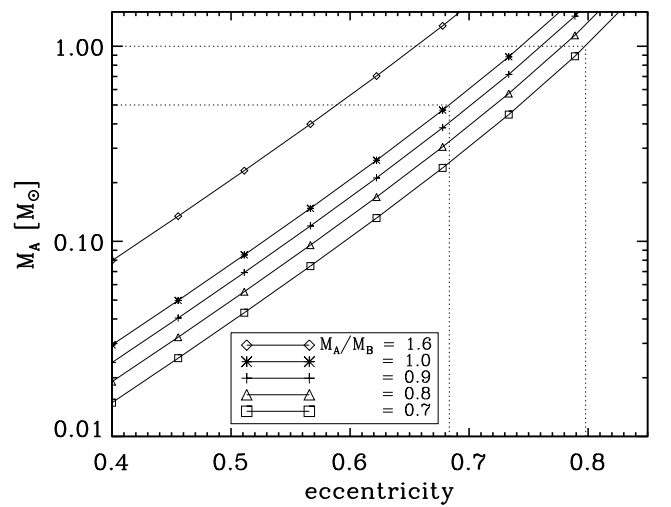}
\caption {The mass of the visible component of KH~15D, $M_A$, versus
orbital eccentricity for five choices of the mass ratio $M_A/M_B$. The
dotted lines show for $0.5 < M_A < 1.0$~\msun and $M_A/M_B \lesssim 1.0$, 
acceptable values of the eccentricity fall within the range 
$0.68 \le e \le 0.80$. The inclination is assumed to be $i =
90^{\circ}$ ($\sin^3 i = 0$).}
\label{mvse}
\end{figure}

\noindent The mass function $F_M/\sin^3i$ in Eqn. (\ref{ma}) is
calculated from the parameters of the 
best-fit orbit to the radial velocity data with the value of $e$ fixed
in the fitting procedure. Figure \ref{mvse} shows plots of $M_A$ versus
$e$ for $M_A/M_B = $ 0.7, 0.8, 0.9 and 1.0. Also shown is the 
value of $M_A/M_B = 1.6$ obtained by W04 based on their model fit to
the \kh light curve (see \S \ref{discussion}). Since the mass of the visible
star is known to fall within the limits $0.5 \le M_A \le 1.0$ M$_\odot$
\citep{flaccomio99,park00}, it can be seen by inspection of Figure
\ref{mvse} that acceptable choices of the eccentricity fall within the
range $0.68 \le e \le 0.80$.

\begin{deluxetable*}{lccc}[!t]
\tablecaption{Optimized Model Parameters \label{parameters}}
\tablewidth{0pt}
\tablehead{
\colhead{Parameter} &
\colhead{Acceptable Fit to} &
\colhead{Acceptable Fit With} &
\colhead{W04}\\
\colhead{} &
\colhead{Velocities\tablenotemark{a}} &
\colhead{$e = 0.74$} &
\colhead{Model 1}
}
\startdata
P~[day]                            & $48.38 \le P \le 48.40$ & \pn (0.01) & 48.35 (fixed)\\
e                                   & $\ge \en$   & 0.74 (fixed) & 0.7 \\
$\omega$~[deg]                      & $-20 \le \omega \le 1$    & -2 (1) & -7.2 \\
$\gamma$\tablenotemark{b}~[\kms]    & $+\vrn \le \gamma \le +22.5$ & +14.7 (0.9) & +15.5\\
$a \sin i$~[AU]                     & $\le 0.29$  & 0.21 (0.5) & 0.18   \\
$K$~[\kms]                          & $\ge 8$   & 53 (4) & 27.5    \\
$F_M/\sin^3 i$       & $2.4\times 10^{-3} \le F_M/\sin^3 i \le 2.2$ & 0.24 (0.05) & $3.8\times 10^{-2}$ \\
T$_p$\tablenotemark{c}~[J.D.]       & --  & 2452251.0 (0.6)& --\\  
Fit rms~[\kms]                      & \rms        & \rms & -- \\
Reduced $\sqrt{\chi_\nu^2}$         &  \csn       & \csn & --          
\enddata
\tablenotetext{a}{The limits on the fit parameters correspond
to the 97.5\% confidence upper limit on $\sqrt{\chi_\nu^2}$ (see
Figure \ref{versuse}) and the condition $e < 1$.}
\tablenotetext{b}{Radial velocity of the binary C.O.M. with respect to the
Solar System barycenter}
\tablenotetext{c}{Time of periastron passage}
\end{deluxetable*}

The resulting range of acceptable orbit parameter values is shown in
Figure \ref{versuse} between the solid, vertical lines in each panel. 
Column two of Table \ref{parameters} lists the best-fit parameter values
obtained using the mean acceptable value of the eccentricity, $e =
0.74$ (see also Figure \ref{ThePlot}). The uncertainties in the fit
parameters were estimated using a Monte Carlo simulation. We generated
$10^3$ statistical realizations of the velocity time series assuming
the errors are normally distributed with standard deviations equal to
the measurement uncertainties. The estimated uncertainties are also
displayed in Figure \ref{versuse} as error bars on the points located
at $e = 0.74$.

\subsection{Periodicity}

As discussed in \S \ref{orbit}, we find the best-fit Keplerian
has an orbital period of $P = \pn \pm 0.01$ days. He02 report 
a period of $P = 48.35 \pm 0.02$ days from their
photometric monitoring and
JW04 report $P=48.42 \pm 0.02$ days from a periodogram
analysis of archival photographic plates photometry. Thus, our
measured orbital period differs by $1.5\sigma$ and 
$2\sigma$ from the modern and historical photometric periods,
respectively. Since the evolving shape of the light curve may complicate
the accurate determination of the photometric period, we conclude that
the orbital period and photometric period of \kh are indistinguishable
within measurement errors.

He02 noted that the shape of the light curve varies from
eclipse to eclipse in such a way as to suggest a 96-day period
for the system, with each cycle containing two brightness
minima. In order to investigate this alternative periodicity, we fixed
the period in our fit at $P = 96.76$ days and obtained a reduced
$\sqrt{\chi_\nu^2}$ of 47.5. Based on this poor quality 
of fit, we find that a 96-day period is incompatible
with our radial velocity measurements. 

\section{Discussion}
\label{discussion}

The results of our spectroscopic monitoring campaign show that
\kh is a spectroscopic binary system. The
observed radial velocities are consonant with a
stellar companion with an orbital period equal to the photometric
period. We now discuss the implications of the binarity of \kh for
current models of the photometric variability mechanism. 

Existing models of the \kh eclipse mechanism fall into two classes
differentiated by whether it is the orbital motion of the star
or a feature of a circumstellar disk that causes the photometric
variability.  
The first class of models posits the existence of a single star
surrounded by a circumstellar disk containing a nonaxisymmetric
density enhancement or alternatively, a warp. As the disk feature
orbits the star with a 
48-day period, it periodically blocks the line-of-sight to the stellar
surface. Based on their spectropolarimetric observations that showed an
increase in polarization during minimum light, \citet{agol04} developed 
a model of a warped disk with an extended atmosphere and obtained a
reasonable fit to the 2001-2002 light curve. A similar 
analysis was performed by \citet{barge03} using a model involving a
large dusty vortex. 

While these single--star models are able to match the light curve
of \kh at one point in time, they do not adequately explain the
temporal changes in 
the observed light curve. Perhaps more importantly, it is not clear if
a binary
companion to \kh is compatible with an eclipsing disk feature. If a
density enhancement in a circumstellar disk is responsible for the 48--day
photometric period, the feature must orbit the
central star at a distance of $\sim 0.22$~AU, assuming a Keplerian
orbit. This geometry would restrict the stellar companion to 
orbit either at a distance less than 0.22 AU or beyond the outer extent
of the circumstellar disk. However, since our best-fit orbit yields a
period of $\pn \pm 0.01$ days and $a\sin i = 0.21$ AU (assuming $e =
0.74$) for the
binary companion, neither of these scenarios seems plausible. 
Thus, it does not seem possible for both an eclipsing disk feature
and a stellar-mass companion to coexist with the same orbital
period. 

In the second class of models, proposed by W04 and CM04,
there exists an unseen binary companion of comparable mass to
the visible K-type star seen today. Surrounding the two stars
is a circumbinary disk viewed nearly edge-on. The orbital plane of the
two stars is tilted at a small angle with respect to the disk plane
and the present day light curve is produced as the reflex motion of Star
A carries it above and below the disk plane.

Independent evidence for the binary nature of \kh
has emerged from studies of archival photographic
plates. In a study of photographic plates obtained from Asiago
Observatory, JW04 show 
that the apparent magnitude of \kh was variable from 1968 to
1983, but the light curve from 
this epoch was markedly different from the one observed today. The
bright state was nearly a factor of 2 brighter in the past and
the eclipse depth was a factor of 5 shallower. These findings
can be explained by invoking the presence of a second star that was
visible in the past but is unseen today (JW04). 

Motivated by these findings, W04 constructed a model of an
eccentric binary with a fraction of the orbital plane obscured by an opaque
screen (or circumbinary disk) and found a quantitative orbital
solution by fitting 
simultaneously to the 2001-2002 photometry of He02, the historical
photometry of JW04 and the radial velocity measurements
of Ha03. Based on a preliminary investigation of the radial velocities
presented here, CM04 independently used physical arguments to propose
a similar model of an eccentric binary surrounded by a nodally
precessing circumbinary ring. As the 
ring precesses, the light curve gradually changes from the one
recovered from the archival plates to the shape seen today.
Perhaps the greatest advantage of the two-star models is their
ability to explain not only the present-day light curve, but also its
evolution over the past half century.  

In addition to explaining the photometric phenomenology of KH~15D, the
two-star models make predictions about the nature of the binary
orbit. Both CM04 and W04 predict that (1) periastron passage occurs during
minimum light, (2) the orbital companion has a mass comparable to the
currently visible star, and (3) the binary orbit is highly eccentric. These
predictions are precisely what we find from our orbit solution. 

The W04 model makes additional, quantitative predictions about the orbital
parameters of the binary. For an assumed fixed period of 48.35 days,
the model produces a velocity semi-amplitude of 27.5~\kms,
eccentricity $e = 0.7$, a mass ratio $M_A/M_B =
1.6$, argument of pericenter $\omega = -7^\circ .2$ and a center of
mass radial velocity $\gamma = +15.5$~\kms. As can be seen in Table
\ref{parameters}, the W04 Model 1 predictions agree well with our
orbital solution with fixed eccentricity, $e = 0.74$ (see \S
\ref{constrain}). We note here that our center of mass velocity for
the binary system of $\gamma = +14.7 \pm 0.9$\kms~clearly rules out
W04 Model 2, which predicts $\gamma = +5.7$~kms and negative velocity
at periapse.

The W04 model also predicts the inclination of the binary orbit, a
property of the system that 
our radial velocities do not. W04 Model 1 predicts $i =
84^\circ.6$ or $\sin i = 0.996$. Similarly, using the geometry of the
circumbinary ring proposed by CM04, an upper limit on the inclination
of $i < 80^\circ$ can be assigned to the binary orbit based on the time lag 
between periastron passage and mid-eclipse (Chiang, private communication).

We note that the W04 model has the peculiar feature that the
less massive star is the more luminous star.
Using Equation \ref{ma}, the values of $P$, $K$ and $e$ produced by
the W04 model lead to a mass function $F_M = 0.038$~\msun~and 
$M_A = 0.41$~\msun. This mass is significantly less than the lower
limit of 0.5~\msun~measured by \citet{park00} and the value of
0.6~\msun~measured by \citet{flaccomio99}. Figure \ref{mvse} shows a
plot of $M_A$ as a function of $e$ for the W04 mass ratio $M_B/M_A =
1.6$. For our 
best-fitting model parameters, only eccentricities between 0.58 and 0.65
yield a mass of A between 0.5 and 1.0 \msun. Therefore,
for a mass ratio of 1.6, the eccentricity reported by W04 ($e = 0.7$)
is larger than the value allowed by our radial velocity measurements
assuming $0.5 \le M_A \le 1.0$~\msun. However, this discrepancy is not too
surprising since the W04 model used only two radial
velocity measurements. It is also important to note
that the mass limits of star A are derived by placing \kh on a
theoretical HR diagram and are therefore subject to the accuracy of
the PMS evolutionary model employed. In all other features of the
\kh binary system, there is a remarkable agreement between the W04
model and the orbital solution calculated from the radial velocities.
Because of the strong evidence of a second star from our
radial velocity measurements and the historical photometry, we find
the two-star class of model to be the most compelling explanation of the \kh
photometric variability mechanism.  

However, the case of the ``winking star'' is still far from closed.
A key missing aspect of the two-star models is direct
detection of a circumbinary disk around KH~15D. He02 report a lack of
near-IR excess and a null detection at millimeter
wavelengths. CM04 state that such findings are consistent with a
circumbinary ring having an inner radius of $\sim1$~AU that is
tidally truncated by the central binary, and an outer radius of
$\sim5$~AU that is possibly shepherded by an as yet unseen
planet. They predict mid-infrared fluxes that are observable with the
Spitzer Space Telescope. Clearly such observations will be vital in
further development of models of the KH~15D system.

\acknowledgements 
We would like to thank Gibor Basri, Paul Butler, Debra Fischer and
Subanjoy Mohanty for generously 
lending portions of their observing time for our project. Many thanks
to Eugene Chiang, Ruth Murray-Clay, Steve Dawson and Josh Winn for
their helpful conversations and suggestions. We acknowledge support by
NASA grant NAG 5-8299 and NSF grant AST95-20443 (to G. W. M.), NASA
grant NAG5-12502 (to W. H.), and
Sun Microsystems. We thank the NASA, University of California and
McDonald Observatory Telescope assignment committees for allocations
of telescope time.

%\bibliographystyle{apj}
%\bibliography{apj-jour,myrefs}

\end{document}